\documentclass{ws-mpla}
\usepackage[super]{cite}
\usepackage{graphicx}
\usepackage{epsfig}
\usepackage{amsmath,amssymb,amsfonts}%
\usepackage{xcolor}

\begin{document}

\markboth{Orhan Donmez}{The comparison of alternative spacetimes using the spherical accretion around the black hole}

\title{The comparison of alternative spacetimes using the spherical accretion around the black hole}


\author{Orhan Donmez}

\address{College of Engineering and Technology, American University of the Middle East, \\
  Egaila 54200, Kuwait.\\
orhan.donmez@aum.edu.kw}

\maketitle

\pub{Received (Day Month Year)}{Revised (Day Month Year)}

\begin{abstract}
  In the region where the gravitational field is strong, we have examined the influence of
  different gravities on the accretion disk formed due to spherical accretion. To achieve this,
  we obtain numerical solutions of the GRH equations, utilizing Schwarzschild, Kerr,
  Einstein-Gauss-Bonnet, and Hartle-Thorne spacetime metrics. We investigate the impact of
  the rotation parameter of a black hole ($a/M$), the EGB coupling constant ($\alpha$), and
  the quadrupole moment of the rotating black hole ($q$) on the accretion disk formed in a
  strong field. The formation of the disk for the slowly and rapidly rotating black hole models
  is separately examined, and comparisons are made. Our numerical simulations reveal that,
  under the  specific conditions, the solution derived from Hartle-Thorne gravity converges
  towards solutions obtained from Kerr and other gravitational models. In the context of the
  slowly rotating black hole with $a/M=0.28$, we observe a favorable agreement between the
  Hartle-Thorne result and the Kerr result within the range of $0 < q < 0.5$. Conversely,
  in the scenario of the  rapidly rotating black hole, a more pronounced alignment with the
  value of $q=1$ is evident within the range of $0.5 < q < 1$. Nevertheless,
  for $q > 1$, it becomes apparent that the Hartle-Thorne solution diverges from solutions
  provided by all gravitational models. Our motivation here is to utilize the Hartle-Thorne
  spacetime metric for the first time in the numerical solutions of the GRH equations for
  the black holes, compare the results with those obtained using other gravities, and
  identify under which conditions the Hartle-Thorne solution is compatible with known black
  hole spacetime metric solutions. This may allow us to provide an alternative perspective
  in explaining observed $X-$ray data.

\end{abstract}

\ccode{PACS Nos.: include PACS Nos.}


\section{Introduction}
\label{Introduction}

The observation and analysis of $X-$ray emissions are crucial steps in explaining astrophysical
systems and their behaviors. For instance, uncovering the physical characteristics of known
and observed compact objects in the universe, understanding the interactions with accretion
disks around them, and finding the physical properties of jets formed as a result of the
interaction between the compact object and the disk are significant strides in unraveling
the mysteries of the universe. Thus, known $X-$ray binaries and active galactic nuclei (AGN)
can be fully understood \cite{Camenzind2006S&W,Tauris2010, Netzer2015ARA&A, Combes}.

Understanding the accretion disk around compact objects plays a crucial role in unraveling the
mysteries of the universe. The accretion disk serves as a laboratory environment for astrophysical
systems, contributing to the explanation of many events occurring in the universe
\cite{Abramowicz2013, Montesinos2012arXiv1203}. Through this disk,
it is possible to comprehend various phenomena, from the formation of celestial objects to
understanding the behavior of matter in environments where strong gravitational forces are present
\cite{Lindoi:10.1146, Inoue2022PASJ}.
In the accretion disk around black holes, matter interacting with the black hole converts a
certain amount of substance into energy, leading to emission. These emissions are observable,
and spectral analyses of these radiations can reveal the characteristics of the compact
object at the center \cite{Luna2018A&A}.

After the gravitational collapse, stable black holes can form, and around these holes, they
can create accretion disks \cite{Jaryal1}. Black holes at the center are characterized by their mass and
angular momentum. The mathematical definition of this black hole was provided by Kerr by solving
Einstein's equations \cite{Misner1973, Schutz2009}. Later, Gauss-Bonnet
gravity was defined as another solution to Einstein's
equations \cite{Glavan2020PhRvL, Errehymy:2023Fortsch, Maurya2023EPJC}. In this gravity, the mass of the
central black hole is determined based on its
angular momentum and the Gauss-Bonnet coupling constant. Using both gravities, it has been observed
that a stable disk is formed around the newly created black hole as a result of
spherical accretion \cite{Donmez2023, Donmez2023arXiv231013847D}.
In this article, in addition to the spacetime matrices mentioned above, we will examine the
behavior of an accretion disk around a slowly rotating compact object, using the Hartle-Thorne
spacetime, which is believed to behave like Kerr and EGB gravities \cite{Kurmanow2023} .
We will also compare the results of these three gravities.

The Hartle-Thorne metric is a solution to Einstein's equations and describes the spacetime
around a slowly rotating compact object. This spacetime matrix defines both the object itself
and the surrounding curvature. Using this spacetime matrix, astrophysical events related to
the rotation of compact objects can be characterized \cite{Kurmanow2023,Destounis2023GReGr}.
Thus, the structure of accretion disks
formed around slowly rotating but deformed black holes can be revealed. The Hartle-Thorne spacetime
varies based on the object's mass, angular, and quadrupole moments. These parameters characterize
different astrophysical systems \cite{Andersson2001CQGra, Stergioulas2003LRR, Stuchl2015AcA}. 
When characterized using the quadrupole moment expression, the Hartle-Thorne metric transforms
into the Kerr metric.

  The Hartle-Thorne metric is widely used in the literature to describe the structure of slowly rotating neutron stars and the phenomena occurring around them. Ref.~\refcite{Stuchl2015AcA} has worked on determining the equations of state compatible with the observed frequency pairs from the source 4U 1636–53 using the Hartle-Thorne geometry. In doing this, the Resonant Switch  model developed for these twin frequency states has been utilized. Thus, based on the compatibility of the Hartle-Thorne metric and the equations of state, the mass, rotation parameters, and quadrupole parameter describing the neutron star and its surrounding geometry for the source 4U 1636–53 have been determined.
The conversion of the Hartle-Thorne metric into the Kerr metric is also crucial for understanding the phenomena surrounding rotating black holes. This transformation is applicable in modeling accretion disks around black holes and in comprehending their diverse behaviors. The dynamic structure of the shock cone formed by Bondi-Hoyle-Lyttleton accretion around the Hartle-Thorne black hole has also been revealed in cases with both slowly and rapidly rotating black holes. Then, the quasi-Periodic-Oscillations trapped within this shock cone were extracted. According to the results obtained in Ref.~\refcite{Donmez2024Univ}, the Hartle-Thorne black hole has the potential to explain the QPOs observed from some sources. Therefore, in addition to known gravitational theories, it can be used to explain physical phenomena and QPO oscillations observed in binary black hole systems.

The application of the Hartle–Thorne metric to systems with a rotating central black hole may
be crucial in explaining certain physical events \cite{Urbancov2019ApJ}.
Alternatively, in some systems where we assume
a central black hole, it may reveal the presence of slowly rotating stars at the center.
Addressing these questions is an important area of research. For this purpose, predictions
made through numerical calculations are significant alongside theoretical models. It has been
observed that the results obtained from numerical models for the $a/M=0.5$ value of the
neutron star are consistent with the Hartle–Thorne metric \cite{Urbancov2019ApJ}.
This underscores the importance of
understanding the disk structures around black holes and comparing them with results obtained
from Kerr and EGB black holes.

In this study, we aim to model accretion disks evolving around Schwarzschild, Kerr, EGB,
and Hartle-Thorne black holes, generated as a consequence of spherical accretion.
In this novel approach, we aim to simulate the accretion disk encircling rotating black holes
by employing the Hartle-Thorne spacetime metric. By doing so, we anticipate uncovering distinct
gravitational effects on the disk around black holes through the application of the Hartle-Thorne
metric, which was initially designed for slowly rotating neutron stars. This undertaking has the
potential to yield fresh perspectives and a more profound comprehension of the astrophysical
phenomena unfolding near rotating black holes, thereby enhancing our understanding of general
relativity in highly gravitational settings. Our
investigation takes into account physical variables such as mass accretion rates, variations in disk
density, the radial velocity of the infalling matter, and other physical
properties of the disk, considering both slow and fast-rotating
black holes. Subsequently, we will compare the results of Schwarzschild, Kerr, and EGB black
holes with each other, as well as compare the results obtained from the Hartle-Thorne
solution with those obtained from these models. Lastly, through
a comparison of numerical results with analytical solutions, we seek to elucidate certain
astrophysical phenomena. To achieve this, we numerically model the GRH equations
using four different spacetime matrices.

After providing matrices that define Hartle–Thorne, Kerr, and EGB black holes and the curvature of
spacetime around them in Chapter \ref{Equations}, we briefly defined the General Relativistic
hydrodynamic equations on the equatorial plane. In Chapter \ref{GRHE2}, we described the
physical characteristics of matter sent from the outer boundary of the computational domain
toward the black hole, forming a global accretion disk around the black hole. Additionally,
in this chapter, we defined the outflow boundary condition that allows matter reaching the
inner boundary to fall into the black hole. Furthermore, in this chapter, we demonstrated
how the horizon of the black hole changes depending on the rotation parameter $a/M$
and the quadrupole moment $q$ in the case of Hartle–Thorne gravity. In Chapter \ref{Results},
we extensively discussed the numerical results of the accretion disk formed around
slowly and rapidly rotating black holes in different gravity scenarios. In the same chapter,
we compared Hartle–Thorne gravity with Kerr gravity, discussing at which value of $q$
in Hartle–Thorne approaches to the Kerr gravity.
Finally, in Chapter \ref{Conclusion}, we summarize the obtained numerical results
and compare with the literature.

Throughout the entire paper, unless explicitly stated otherwise, the convention of using
geometrized units where $c$  and $G$ are set to $1$ is adhered to.
All the parameters used in this article are dimensionless. They all vary proportionally with the
mass of the central black hole. Therefore, the results obtained here can be applied to stellar
or massive black holes. Only by using the mass of the black hole at the center of the applied
system, a conversion is made from geometricized units to the SI unit system. Subsequently,
the numerical results obtained can be compared with observational outcomes.


\section{Equations}
\label{Equations}

\subsection{Hartle-Thorne Spacetime Metric}
\label{Hartle_Thorne}

The Hartle-Thorne spacetime represents a solution to Einstein's field equations in vacuum,
providing a description of the curved spacetime surrounding a rotating black hole. This metric
accurately accounts for the mass quadrupole moment $(q=Q/M^3)$ to first order
and the angular momentum $(a=J/M^2)$
to second order where $M$ is the mass of the black hole.
The spacetime metric pertains to a slowly rotating black hole with nearly
spherical characteristics. Utilizing this metric in General Relativity and Hydrodynamics
equations offers a valuable avenue to investigate the behavior of matter and understand
the accretion mechanisms occurring in the strong gravitational region near the event horizon
of the rotating black hole \cite{Donmez2024Univ}.

The line element of the Hartle-Thorne spacetime reads \cite{Kurmanow2023, Cromartie2020, Donmez2024Univ}

\begin{eqnarray}
  ds^2 = -\left( 1- \frac{2M}{r}\right)\left(1 + 2f_1 P_2\left(Cos\theta \right) +
  2\left( 1- \frac{2M}{r}\right)^{-1}\frac{J^2}{r^4}\left(2Cos^2\theta -1\right) \right)dt^2 \nonumber \\
 + \left( 1- \frac{2M}{r}\right)^{-1}\left(1 - 2f_2 P_2\left(Cos\theta \right) -
 2\left( 1- \frac{2M}{r}\right)^{-1}\frac{J^2}{r^4} \right)dr^2  \\
 -\frac{4J}{r}Sin^2\theta dtd\phi + r^2\left[1-2f_3 P_2\left(Cos\theta \right)\right] d\theta^2
 r^2\left[1-2f_3 P_2\left(Cos\theta \right)\right] Sin^2\theta d\phi^2,   \nonumber
\label{HT1}
\end{eqnarray}

\noindent
where
$P_2\left(Cos\theta \right) = \frac{1}{2}\left(3Cos^2\theta - 1\right)$, \;
$f_1 = \frac{J^2}{M r^3} \left(1+\frac{M}{r}\right) + \frac{5}{8}\frac{qM^3 -J^2/M}{M^3}Q_2^2 \left(\frac{r}{M}-1\right)$,
\; $f_2 = f_1 - \frac{6J^2}{r^4}$,
\;$f_3 = f_1 + \frac{J^2}{r^4} + \frac{5}{4}\frac{qM^3 -J^2/M}{M^2\left(r^2-2Mr\right)^{1/2}}Q_2^1 \left(\frac{r}{M}-1\right)$. 
$P_2\left(Cos\theta \right)$ is the second Legendre polynomial of the first kind.  $Q_2^1$ and  $Q_2^2$ are the associated
Legendre Polynomials of the second kind which are
$Q_2^1 = \left(x^2-1\right)^{1/2}\left(\frac{3x}{2}ln\frac{x+1}{x-1}-\frac{3x^2-2}{x^2-1}\right)$ and 
$Q_2^2 = \left(x^2-1\right)\left(\frac{3}{2}ln\frac{x+1}{x-1}-\frac{3x^3-5x}{x^2-1}^2\right)$ as a
function of $x$ which is $x=\frac{r}{M}-1$.

In order to solve the GRH equations using the spacetime matrix, we need to define the lapse
function and shift vectors in the Hartle-Thorne coordinates. The relation between $4$-metric $g_{ab}$,  
$3$-metric $\gamma_{ij}$, lapse function and shift vectors given as \cite{Misner1977}

\begin{eqnarray}
  \left( {\begin{array}{cc}
   g_{tt} & g_{ti} \\
   g_{it} & \gamma_{ij} \\
  \end{array} } \right)
=  
  \left( {\begin{array}{cc}
   (\beta_k\beta^k - \alpha^2) & \beta_k \\
   \beta_i & \gamma_{ij} \\
  \end{array} } \right),
\label{HT2}
\end{eqnarray}

\noindent
where $i, j, k= 1, 2$ and $3$. After doing straightforward calculation, the lapse function for Hartle-Thorne
spacetime metric is

\begin{eqnarray}
\alpha = \sqrt{\frac{4J^2}{r^4(1+f_3)} - \left(1-f_1 - \frac{2J^2}{r^4\left(1-\frac{2M}{r}\right)}\right)\left(\frac{2M}{r}-1 \right)},
\label{HT3}
\end{eqnarray}

\noindent
the components of the shift vectors are 

\begin{eqnarray}
  \beta_r = 0, \;\;\;  \beta_{\theta}= 0, \;\;\;  \beta_{\phi}=-\frac{2J}{r}
\label{HT4}
\end{eqnarray}

As previously mentioned, the external Hartle-Thorne geometry can be described using three parameters,
namely the black hole mass (M), the dimensionless angular momentum $(a)$, and the dimensionless quadrupole moment $(q)$.


\subsection{Kerr and EGB Spacetime Matrices}
\label{Kerr_EGB}

One of the general objectives of this article is to compare the results obtained for the accretion
disks modeled using the Hartle-Thorne geometry  with the results previously found in the Kerr and
Einstein-Gauss-Bonnet (EGB) gravity scenarios. Therefore, briefly, the spacetime matrices for
Kerr and EGB gravities are provided below with  supporting references.
Utilizing various matrices in numerical computations further enables us to investigate the
implications of alternative gravitational theories, as demonstrated in previous studies
\cite{Donmez2023arXiv231013847D, Ghosh2, Donmez3, Donmez_EGB_Rot, Donmez2024arXiv}.

The Kerr black hole exhibits an exceptionally potent gravitational force owing to its rotational
parameter, simultaneously distorting the spacetime in its vicinity. Specifically, when
investigating the characteristics of material within the inner disk region, the Kerr metric
is the appropriate choice. The Kerr black hole's metric in Boyer-Lindquist coordinates is
defined as follows\cite{Misner1973, Schutz2009,Donmez6}:

\begin{eqnarray}
  ds^2 = -\left(1-\frac{2Mr}{\sum^2}\right)dt^2 - \frac{4Mra}{\sum^2}sin^2\theta dt d\phi
  + \frac{\sum^2}{\Delta_1}dr^2 + \sum^2 d\theta^2 + \frac{A}{\sum^2}sin^2\theta d\phi^2,
\label{Kerr1}
\end{eqnarray}

\noindent where $\Delta_1 = r^2 - 2Mr +a^2$, and $A = (r^2 + a^2)^2 - a^2\Delta sin^2\theta$.
In Boyer-Lindquist coordinates, the shift vector and lapse function of the Kerr metric are given by:
Shift vector ($\beta^i$): $\beta^t = 0$, $\beta^r = 0$, and $\beta^\theta = \frac{-2Mar}{A}$.
The lapse function  is  $\tilde{\alpha} = \sqrt{\frac{\Delta_1}{A}}$.

Einstein-Gauss-Bonnet (EGB) gravity is a distinct solution within the framework of Einstein's
equations, offering an alternative gravitational theory. Unlike Kerr gravity, this alternative
theory features the presence of an alpha ($\alpha$) parameter. The influence of this parameter,
particularly when investigated in proximity to black holes, can be harnessed to provide
explanations for observed phenomena. The $\alpha$ parameter introduces higher curvature
corrections, thereby providing a novel viewpoint on the behavior of black holes under
extreme conditions and within higher-dimensional spacetime \cite{Fernandes2022}. EGB metric is
\cite{Glavan2020PhRvL, Donmez_EGB_Rot, Donmezetal2022}

\begin{eqnarray}
  ds^2 &=& -\frac{\Delta_2 - a^2sin^2\theta}{\Sigma}dt^2 + \frac{\Sigma}{\Delta_2}dr^2 -
  2asin^2\theta\left(1- \frac{\Delta_2 - a^2sin^2\theta}{\Sigma}\right)dtd\phi + 
  \Sigma d\theta^2 + \nonumber \\
  && sin^2\theta\left[\Sigma +  a^2sin^2\theta \left(2- \frac{\Delta_2 -  
   a^2sin^2\theta}{\Sigma} \right)  \right]d\phi^2,
\label{GRHEq6}
\end{eqnarray}

\noindent
where $\Sigma = r^2 + a^2 \cos^2\theta$,
$\Delta_2 = r^2 + a^2 + \frac{r^4}{2\alpha}\left(1 - \sqrt{1 +\frac{8 \alpha M}{r^3}} \right)$.
In this context, $a$, $\alpha$, and $M$ correspond to the black hole's spin parameter,
Gauss-Bonnet coupling constant, and its mass, respectively.
The lapse function is $\tilde{\alpha} = \sqrt{\frac{a^2(1-f(r))^2}{r^2+a^2(2-f(r))} + f(r)}$
and the shift vectors are
$\beta^i = \left(0,\frac{a r^2}{2\pi \alpha}\left(1 - \sqrt{1 + \frac{8 \pi\alpha M}{r^3}}\right),0\right)$.
The variable $f(r)$ is defined as
$f(r) = 1 +\frac{r^2}{2\alpha}\left(1 - \sqrt{1 + \frac{8 \alpha M}{r^3}} \right)$. 
The gamma matrix, denoted as $\gamma_{i,j}$ in the GRH equations, defines three-dimensional
space and is derived from the metrics $g_{ab}$ for both Kerr and EGB gravities.
In this context, the Latin indices $i$ and $j$ vary from $1$ to $3$.


\subsection{General Relativistic Hydrodynamic Equations}
\label{GRHE1}

During gravitational collapse, a fluid, such as gas or dust, is drawn toward a massive
object, such as a black hole, neutron star or massive star, by gravitational forces and
accumulates around it. This phenomenon plays a crucial role in comprehending the interaction
between matter and  black hole, as well as other dense entities in the universe.
The examination of the gravitational collapse of a perfect fluid in the vicinity of
black holes, particularly the Kerr, Einstein-Gauss-Bonnet (EGB), and Hartle-Thorne black holes,
involves solving General Relativistic Hydrodynamical (GRH) equations within a curved background.
The perfect fluid stress-energy-momentum tensor is given as

\begin{eqnarray}
 T^{ab} = \rho h u^{a}u^{b} + P g^{ab},
\label{GREq1}
\end{eqnarray}

\noindent $g^{ab}$, $\rho$, $u^{a}$, $p$, and $h$,  are  the three-metric of the curved spacetime
the rest-mass density, the $4-$ velocity of the fluid, the fluid pressure,
and the specific enthalpy, respectively.
The indices $a$, $b$, and $c$ range from $0$ to $3$.

\noindent In order to numerically solve the GRH equations, it is essential to represent
them in a conserved form \cite{Donmez1}:

\begin{eqnarray}
  \frac{\partial U}{\partial t} + \frac{\partial F^r}{\partial r} + \frac{\partial F^{\phi}}{\partial \phi}
  = S.
\label{GREq4}
\end{eqnarray}

\noindent The vectors $U$, $F^r$, $F^{\phi}$, and $S$ correspond to the conserved variables,
fluxes along the $r$ and $\phi$ directions, and sources, respectively. These conserved
variables are defined in relation to the primitive variables, as illustrated below,

\begin{eqnarray}
  U =
  \begin{pmatrix}
    D \\
    S_r \\
    S_{\phi} \\
    \tau
  \end{pmatrix}
  =
  \begin{pmatrix}
    \sqrt{\gamma}W\rho \\
    \sqrt{\gamma}h\rho W^2 v_r\\
    \sqrt{\gamma}h\rho W^2 v_{\phi}\\
    \sqrt{\gamma}(h\rho W^2 - P - W \rho),
    \end{pmatrix}
\label{GREq5}
\end{eqnarray}

\noindent  where the term $h = 1 + \epsilon + P/\rho$ represents the enthalpy,
the expression $W = (1 - \gamma_{a,b}v^i v^j)^{-1/2}$ denotes the Lorentz factor,
$v^i = u^i/W + \beta^i$ represents the three-velocity of the fluid.
$\epsilon$ represents the internal energy.
The fluid pressure is calculated using the ideal gas equation of state. The
three-metric $\gamma_{i,j}$ and its determinant $\gamma$ are determined based on the
four-metric of black hole. Latin indices $i$ and $j$ vary from $1$ to $3$.
The flux and source terms can be computed for any metric using the following equations,

\begin{eqnarray}
  \vec{F}^i =
  \begin{pmatrix}
    \tilde{\alpha}\left(v^i - \frac{1}{\tilde{\alpha}\beta^i}\right)D \\
    \tilde{\alpha}\left(\left(v^i - \frac{1}{\tilde{\alpha}\beta^i}\right)S_j + \sqrt{\gamma}P\delta^i_j\right)\\
    \tilde{\alpha}\left(\left(v^i - \frac{1}{\tilde{\alpha}\beta^i}\right)\tau  + \sqrt{\gamma}P v^i\right)
    \end{pmatrix}
\label{GREq6}
\end{eqnarray}

\noindent and,

\begin{eqnarray}
  \vec{S} =
  \begin{pmatrix}
    0 \\
    \tilde{\alpha}\sqrt{\gamma}T^{ab}g_{bc}\Gamma^c_{aj} \\
    \tilde{\alpha}\sqrt{\gamma}\left(T^{a0}\partial_{a}\tilde{\alpha} - \tilde{\alpha}T^{ab}\Gamma^0_{ab}\right)
   \end{pmatrix} 
\label{GREq7}
\end{eqnarray}

\noindent where $\Gamma^c_{ab}$ is the Christoffel symbol.


\section{Initial and Boundary Conditions}
\label{GRHE2}

Here, using the Hartle-Thorne metric, we examine the disk of matter that spreads around the
newly formed compact object due to gravitational collapse, revealing the structures of
this disk. By comparing the obtained results with those from other gravity models such as
Kerr and Gauss-Bonnet, we determine the effects of alternative gravity. To perform these
comparisons, we numerically solve these metrics in the General Relativistic Hydrodynamic (GRH)
equations \cite{Donmez1, Donmez5, Donmez2}. While solving these equations, we assume that
matter behaves like an ideal gas and use $P = (\Gamma - 1)\rho\epsilon$ with $\Gamma = 4/3$.

Initially, we assume that the computational domain around the black hole is empty. Therefore,
during the definition of this region, we consider it to have negligible density and pressure,
and these values should be exceedingly small. While making these choices, density and pressure
values are determined with the consideration that the speed of sound should be
$C_{\infty} = 0.1$. The same selection is applied to the initial values used when matter
is sent into the computational domain from outer boundary.
For the formation of the disk around the black hole, gas from the outer boundary is injected
with values $V^r = -0.01$, $V^{\phi}=0$, and $\rho=1$ \cite{Donmez2023}.
Calculations are performed on the
equatorial plane. Detailed descriptions of the initial models for Kerr, Gauss-Bonnet, and
Hartle-Thorne black hole, can be found in Table \ref{Inital_Con}.
  
\begin{table}
\scriptsize
\caption{The initial model adopted for the numerical simulation of Kerr, Gauss-Bonnet, and
  Hartle-Thorne metric.
  $Model$, $type$, $\alpha$,  $a/M$, and  $q$ are the name of the model,
  the gravity, Gauss-Bonnet coupling constant, the black hole rotation parameter, and
  quadrupole parameter, respectively.}
 \label{Inital_Con}
  \begin{tabular}{ccccc}
    \hline
    \hline

    $Model$        & $type$         & $\alpha (M^2)$ & $a/M$ & $q$ \\
    \hline
    $SCH$          &$Schwarzschild$ & $-$       & $-$    & --     \\
    \hline    
    $K028$          & $Kerr$        & $-$       & $0.28$ & --     \\    
    $K09$          &                & $-$       & $0.9$  & --     \\
    \hline
    $K028\_EGB1$   &                & $0.68$    & $0.28$ & --      \\
    $K028\_EGB2$   &                & $-3.03$   & $0.28$ & --      \\   
    $K028\_EGB3$   & $Gauss-Bonnet$ & $-4.93$   & $0.28$ & --      \\
    $K09\_EGB1$    &                & $0.05$    & $0.9$  & --      \\
    $K09\_EGB2$    &                & $-3.61$   & $0.9$  & --      \\ 
    \hline
    $K028\_HT1$    &                & --        & $0.28$ & $0$      \\
    $K028\_HT2$    &$Hartle-Thorne$ & --        & $0.28$ & $1$      \\
    $K028\_HT3$    &                & --        & $0.28$ & $3$      \\
    $K028\_HT4$    &                & --        & $0.28$ & $5$      \\
    \hline          
    $K09\_HT1$     &                & --        & $0.9$  & $0$      \\
    $K09\_HT2$     &$Hartle-Thorne$ & --        & $0.9$  & $1$      \\
    $K09\_HT3$     &                & --        & $0.9$  & $3$      \\
    $K09\_HT4$     &                & --        & $0.9$  & $6$      \\    
    \hline
    \hline
  \end{tabular}
\end{table}

The grid employs uniformly spaced zones in both the radial and angular directions, with $1024$
zones in the radial direction and $256$ zones in the angular direction.
The inner boundary of the computational domain is situated at $r_{min}=3.7M$,
while the outer boundary extends to $r_{max}=100M$ along the radial axis.
Angular boundaries are defined as $\phi_{min}=0$ and $\phi_{max}=2\pi$.
In this article, we investigate the behavior of matter falling towards a black hole through
accretion in order to reveal the effects of different gravities on the region near the black hole
horizon. Therefore, it is necessary for the inner boundary of the computational domain to be as close to
the black hole horizon as possible. However, the location of the horizon of EGB black holes varies
entirely depending on the  $\alpha$ parameter. For instance, when $\alpha = 0.6$, the horizon is
around $r=1M$, whereas when $\alpha=-4.93$, the horizon is at $r=3.5M$. Consequently,
to ensure a fair comparison, the inner radius of the computational domain is set
to $r=3.7M$ in each model.

The code is executed until the time $(t_{max} = 30000M)$, which significantly surpasses
the time $(\sim 5000M)$ required for the model to reach a steady state. It has been ascertained
that crucial aspects of the numerical solutions, such as instabilities, the presence of
quasi-periodic oscillations (QPOs), the location of shocks, and the behavior of
accretion rates, remain largely unaffected by variations in grid resolution
\cite{Donmez2023arXiv231013847D, Donmez3, Donmez_EGB_Rot,Donmezetal2022}.

The accurate definition of boundaries in numerical modeling prevents unwanted oscillations.
Thus, the characteristics of oscillations arising from black hole-disk interactions can be
revealed. Otherwise, unwanted residues from the boundary can create both non-physical
situations and lead to the crash of the code \cite{Donmez2023}. Here, to model the
proposed disk, we aim for matter near the inner boundary, close to the black hole, to fall
towards the black hole. This is because once matter crosses the last stable orbit, it is
impossible for it to escape from the black hole. Therefore, an outflow boundary condition is
applied here. On the other hand, according to the proposed model, it is assumed that the
supernovae remnants continuously fall towards the black hole. As a result, matter is
injected at the outer boundary of the disk, which is far from the black hole.

\begin{figure*}
  \vspace{1.3cm}
  \center
  \psfig{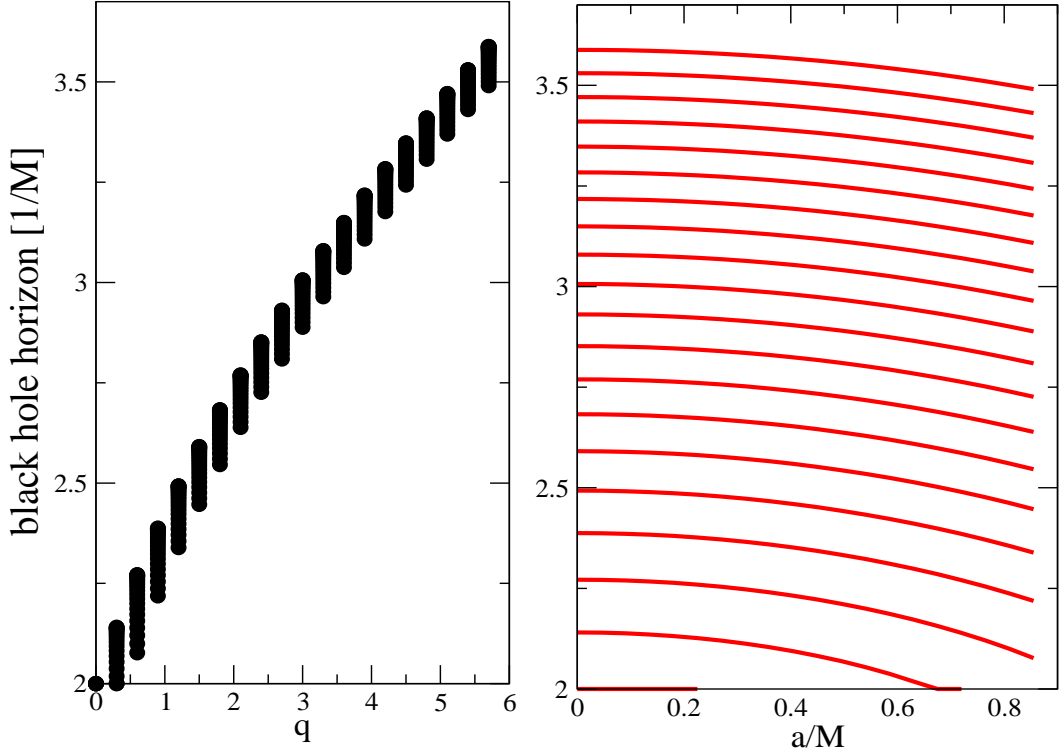}   
  \caption{The change in the black hole's horizon depending on the black hole's spin ($a/M$)
    and quadrupole parameter ($q$). The left panel shows the change in the
    black hole's horizon for each $a/M$ as a function of $q$  while the right one
    illustrates the change in the horizon  for each $q$ as a function of $a/M$.}
\label{star_radius}
\end{figure*}

In Fig.\ref{star_radius}, the variation of the radius of the compact object defined by the
Hartle-Thorne gravity is shown according to $q$  and $a/M$. The left part of the figure
illustrates the change in radius with respect to $q$ in different $a/M$ conditions.
Each thick black vertical line represents the change in radius for a different $a/M$. Here,
it is clearly seen that as $q$ increases, the radius increases. However, the growth of $a/M$,
as in the Kerr spacetime matrix, has reduced the radius. This confirms that the Hartle-Thorne
metric behaves similarly to Kerr \cite{Kurmanow2023}. On the other hand, the right part of the
figure shows how the radius changes with $a/M$. Each point corresponding to the thick lines in
the vertical direction, where $a/M$ is constant, corresponds to different $q$. This indicates
that increasing $q$ increases the radius. Again, it is clearly seen here that increasing $a/M$
reduces the radius. Finally, it has been observed that the value of $a/M$ corresponding to each
$q$ is not between $0$ and $1$. In other words, there is a certain range of physically defined
$a/M$ for each $q$. The reverse is also true. However, it has been observed that as $q$
increases, the number of possible values for $a/M$ increases. In other words, to be able to
choose $a/M$ anywhere between $0$ and $1$, $q$ must be sufficiently large.

\section{The Numerical Simulation of the Spherical Accretion}
\label{Results}

The accretion disks are important mechanisms to describe the physical characteristics of the
black hole at the center. Understanding the properties of the electromagnetic emissions emitted by
the disks is necessary to comprehend the characteristics of black holes. For decades, scientists have
been working to understand the enigmatic creatures of the universe and  black holes by using theoretical,
numerical, and observational methods. Important observations have been attempted to understand
the physical characteristics of the black holes at the centers of the Milky Way and $M87$ galaxies,
especially through significant observations made by the Event Horizon Telescope (EHT)
in recent years \cite{Akiyama1,Akiyama2}.
Particularly, observations indicate that the geometry around black holes has certain
limitations \cite{Abramowicz2013}.
In this context, numerical and theoretical studies are crucial to contribute to observational
results. Many studies have been conducted, especially in revealing the characteristics of accretion
disks around Schwarzschild and Kerr black holes, both theoretically
and numerically \cite{Abramowicz2013,Montesinos2012arXiv1203,Chakrabarti2002, Koyuncu1}.

In recent years, with the definition of 4D EGB gravity \cite{Ghosh2},
theoretical \cite{Feng1,Zhang1,Islam1,Kumar1,Ghosh1,Wei1,Fard1,Haydari2} and
numerical studies \cite{Donmez3,Donmez_EGB_Rot} the
dynamic structures of accretion disks around the $EGB$ black holes and the resulting $QPO$ oscillations
continue. Additionally, adapting Hartle-Thorne gravity, which is an important gravity in describing the behavior
of matter around neutron stars, to black holes and conducting
theoretical \cite{Kurmanow2023, Destounis2023GReGr,Urbancov2019ApJ} and numerical studies
in this direction can provide a different approach to explaining disk characteristics and
observational data.
In this context, the theoretical investigation of the structure of the disk around an Hartle-Thorne
black hole has begun, and the support of this by numerical relativity would be a significant
advancement in the field. This is because Hartle-Thorne gravity is an important theory in explaining
slowly rotating deformed objects in regions where the gravity
is very strong \cite{Kurmanow2023, Destounis2023GReGr}. Thus, by examining
the three crucial parameters characterizing a black hole—mass, angular momentum, and quadrupole moment
in the context of Hartle-Thorne gravity, certain astrophysical system
characteristics can be revealed \cite{Andersson2001CQGra,Stergioulas2003LRR}.

Comparing Hartle-Thorne gravity with Kerr and EGB ones, it has some advantages, especially in terms
of its flexibility \cite{Kurmanow2023}. In the absence of angular momentum, it characterizes a naked singularity.
If there are angular momentum and a quadrupole moment, it behaves like the Kerr metric. For these
reasons and other things mentioned above, the Hartle-Thorne metric applied to the  black holes in
recent years. Modeling the disk around black holes with Hartle-Thorne Horndeski gravity could
open a different perspective in explaining observational data.
By numerically solving the GRH equations using this Hartle-Thorne  metric, we reveal how
the parameters of an accretion disk behave in different gravities in the following sections.

\subsection{Slowly Rotating Black Hole: a/M=0.28}
\label{slowly}

\begin{figure*}
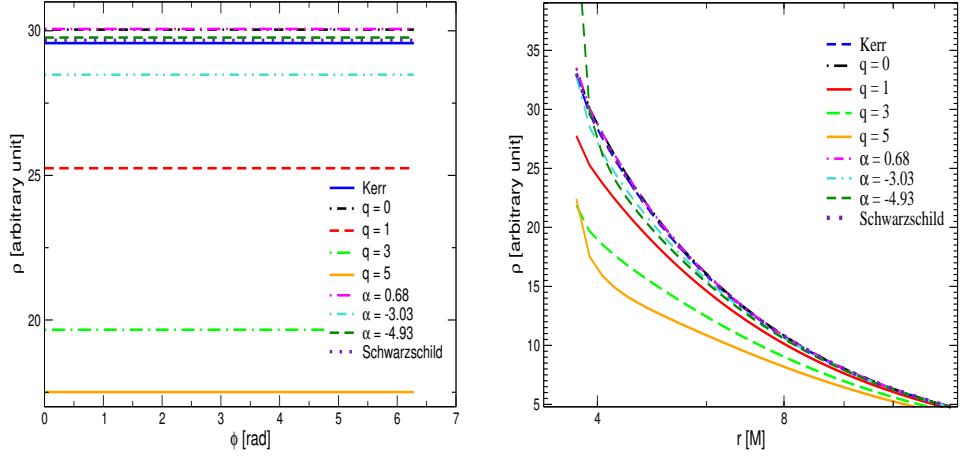

  \vspace{1.4cm}
  \center
  \psfig{file=a028_fixed_r_dens.eps,width=6cm,height=6cm} \hspace{0.4cm}
  \psfig{file=a028_rho.eps,width=6cm,height=6cm}    
  \caption{In the case of a black hole with a spin parameter of $a/M=0.28$, the variation of
    density with respect to position at $t=30000M$. On the left graph, the density variation
    with respect to $\phi$ for the accretion disk is shown, while on the right, it is illustrated
    how this density changes with $r$. In both cases, the density variation is compared for
    different gravities and their respective parameters.}
\label{a028_fixed_r_dens}
\end{figure*}

The rest-mass density resulting from spherical accretion can be used to explain many physical
events occurring due to the interaction of the black hole with its accretion disk. In other words,
information about the mass density can be derived to understand the physical characteristics and
types of electromagnetic radiation observed by detectors resulting from this interaction.
The rest-mass density is also related to the mass accretion rate of the spherical accretion around
the black hole. The density obtained as matter falls towards the black hole
provides information about the amount of matter falling into the black hole. Thus, an estimate can be
made about the mass increase rates of the black holes. For these reasons, numerically calculated
rest-mass density using different gravities provides us with information about the spacetime
curvature around the black hole. This is why, in Fig.\ref{a028_fixed_r_dens},
we showed how the mass density of the
disk around the slowly rotating black holes changes radially and axially using the different gravities.

The left part of  Fig.\ref{a028_fixed_r_dens}
shows how the density changes in the azimuthal direction at $r=4M$ for the
slowly rotating black hole model with $a/M=0.28$. As expected, since the disk is obtained through
global accretion, the density remains constant. However, the density exhibits different behaviors for
different gravities and the parameters associated with those gravities, as anticipated. This,
as mentioned above, contributes to the variations observed in the data and the change in the mass
of the central black hole. In the right panel of Fig.\ref{a028_fixed_r_dens},
the variation of the disk's density with respect to $r$ is shown for different gravities and parameters.
As expected, due to the strong gravitational field, the density of the disk has exponentially
increased as it approaches the black hole  horizon.

In the Fig.\ref{a028_fixed_r_dens}, the variation in mass resulting from the use of Kerr,
Schwarzschild, and other gravities is
provided. Since the black hole is the slowly rotating, the results for Kerr and Schwarzschild are
close to each other. EGB gravity exhibits different behavior for extreme positive and negative values of
$\alpha$, corresponding to this rotation parameter. When $\alpha=0.68$, the mass density is greater
than Kerr, and for negative values of $\alpha$, it is smaller than Kerr. Interestingly, at
$\alpha=-4.93$, it is observed that the density is slightly larger than Kerr again in the
strong gravitational region (very close to the horizon). This is related
to the intensive chaos occurred during the black hole-matter interaction and it was also mentioned in
our previous calculations \cite{Donmez_EGB_Rot, Donmez2023arXiv231013847D,Donmez2023Propose}.
However, as seen in the right panel of Fig.\ref{a028_fixed_r_dens}, it has been observed that
as we move away from the black hole's horizon, this value is still smaller than Kerr.

On the other hand, the behavior of the disk's density in Hartle-Thorne gravity is compared to Kerr
gravity. As seen in the Fig.\ref{a028_fixed_r_dens}, at $q=0$, the maximum density of the disk is
greater than the Kerr case, but at $q=1$, this density is smaller than Kerr. It is clearly
seen that as the $q$ value increases, the density becomes smaller compared to Kerr, indicating
a departure from the Kerr value. In conclusion, as understood from this graph, in Hartle-Thorne
gravity, it can be said that it approaches Kerr for the range of $0 < q < 1$.
More specifically, it can be inferred that it is compatible with Kerr when the $q$ is
slightly less than $0.5$.

\begin{figure*}
  \vspace{1.4cm}
  \center
  \psfig{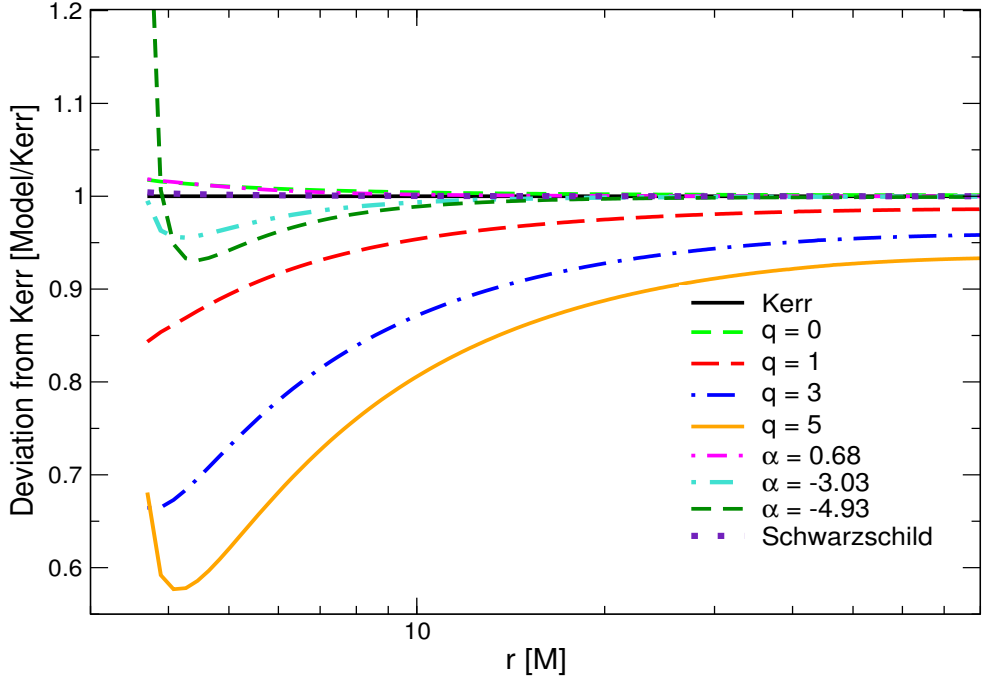}   
  \caption{The same as Fig.\ref{a028_fixed_r_dens}, but this time comparing the radial
    disk density for each gravity after normalization with Kerr solution.
    In the presence of different gravities, the radial variation of the density of the
    accretion disk formed through spherical accretion around a black hole is compared
    with the disk density around a Kerr black hole.}
\label{a028_deviation}
\end{figure*}

To observe how much different gravities deviate from Kerr, or in other words, the deviation
ratios from Kerr, we normalized the results obtained from all models with Kerr. This situation is
presented in Fig.\ref{a028_deviation}. As seen here, in the $q=0$ case, the rest-mass density has
produced a solution closer to Kerr, while this deviation is more significant in the $q=1$ case.
As mentioned above, for a value of $q$ in the range $0 < q < 0.5$, we can say that the
Hartle-Thorne solution transforms to Kerr. Again, as seen in Fig.\ref{a028_deviation}, interestingly,
for large negative $\alpha$ values in EGB gravity, it exhibits similar behavior to Hartle-Thorne
with larger $q$. However, while the deviation of EGB gravity results occurs around Kerr,
for large values of  $q$, the results gradually move away from the Kerr solution.

\begin{figure*}
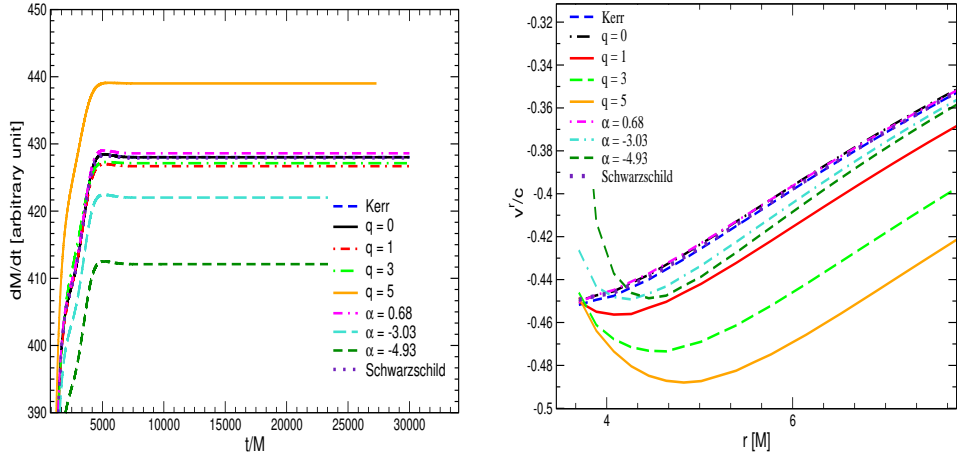

  \vspace{1.4cm}
  \center
  \psfig{file=a028_mass_acc.eps,width=6cm,height=6cm}   \hspace{0.4cm}
  \psfig{file=a028_vr.eps,width=6cm,height=6cm}    
  \caption{The same as Fig.\ref{a028_fixed_r_dens}, but the variation of the mass accretion
    rate of the disk formed around a slowly rotating black hole and the radial variation of the
    infalling matter's radial velocity towards the black hole are provided.}
\label{a028_mass_acc}
\end{figure*}

The mass accretion rate is directly related to the matter density around a black hole. A
higher mass accretion implies that the disk is hotter, leading to the production of high-energy
radiation. On the other hand, a higher mass accretion means more matter falling into the black
hole, resulting in a decrease in the rest-mass density of the disk. Mass accretion, and
consequently the rest-mass density, is associated with the luminosity of the disk. Therefore,
changes in the mass accretion rate in different gravitational scenarios lead to variations in
the observed data. These variations are depicted in the left part of Fig.\ref{a028_mass_acc}.
In Hartle-Thorne gravity, as the $q$ value increases (for example, $q=5$), and in EGB
gravity, for large negative values of $\alpha$ (for example, $\alpha= -3.03$ and $-4.93$),
the mass accretion rate significantly deviates from other models. However, models other than
these exhibit similar mass accretion rates, meaning they can produce similar luminosity.

The graph on the right side of Fig.\ref{a028_mass_acc} illustrates the radial velocity of
matter falling towards the black hole as a function of $r$. This allows the estimation of the
conversion rate from gravitational potential energy, created by the black hole on the matter,
to kinetic energy. Additionally, the radial velocity shows that in the region where the
gravitational force is high, i.e., in the region where the potential well created by spacetime
is deepest, the flow velocity of matter decreases. Again, it is observed that the Hartle-Thorne
result for $0 < q < 0.5$ might be consistent with Kerr solution.

\subsection{Rapidly Spinning Black Hole: a/M=0.9}
\label{rapidly}

Although Hartle-Thorne gravity describes the spacetime around a slowly rotating deformed compact
object, comparing the results obtained from Hartle-Thorne gravity for cases with high rotation
parameters to other gravitational theories and Kerr may help us better understand not only
slowly rotating black hole systems but also provide an opportunity to compare different
gravitational theories in such scenarios. The numerical modeling of accretion disk around
the black hole in  Hartle-Thorne gravity may be utilized to explain certain astrophysical
phenomena in the case of a rapidly rotating black hole. Considering both the rapid rotation
and deformation of the black hole can be crucial in understanding such scenarios.
In this situation, unraveling the behavior of globally accreting matter can contribute significantly
to the literature. On the other hand, the primary goal of Hartle-Thorne gravity, which is actually
to examine the behavior of matter around a deformed compact object rather than the rotation parameter,
can also be explored in the context of the impact of the rotation parameter. Therefore,
in this section, the accretion disk around a rapidly rotating Hartle-Thorne black hole is modeled,
and the results are compared with different gravitational models in the presence of the rapid rotation
parameter.

\begin{figure*}
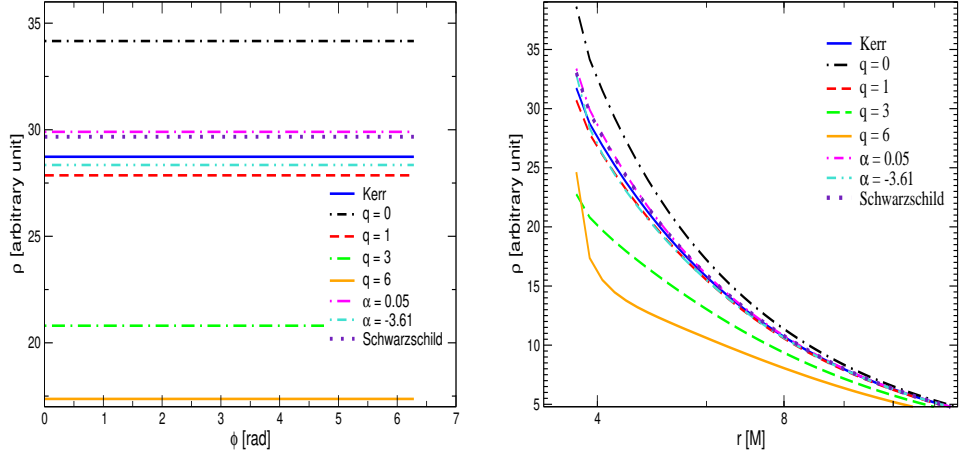

  \vspace{1.4cm}
  \center
  \psfig{file=a09_fixed_r_dens.eps,width=6cm,height=6cm} \hspace{0.4cm}
  \psfig{file=a09_rho.eps,width=6cm,height=6cm}    
  \caption{The same as Fig.\ref{a028_fixed_r_dens} but it is for a spin parameter of $a/M=0.9$.}
\label{a09_fixed_r_dens}
\end{figure*}

We compared different gravities with the Kerr model for the rapidly rotating black hole scenario
with $a/M=0.9$, drawing the rest-mass density in both the azimuthal and radial directions. The
left part of Fig.\ref{a09_fixed_r_dens} illustrates the change in density with respect to $\phi$.
Again, as expected, the disk density remains constant along the  $\phi$ at $r=4M$. However,
it is evident that the rest-mass density takes different values around the Kerr model for
different gravities and their associated parameters. The significant variation in the
rest-mass density clearly indicates that different gravities, and consequently the black holes,
introduce differences in the dynamical structure of the accretion disk, leading to potential
oscillations, radiation, and variations in luminosity.

As seen in Fig.\ref{a09_fixed_r_dens}, within the rapidly rotating black hole model,
the Hartle-Thorne solution exhibits variation around the Kerr solution for the range of $0 < q < 1$
for the quadrupole moment parameter. Upon closer examination, it is observed that the disk's
rest-mass density approaches the Kerr solution as $q$ approaches $1$. Therefore, it is
anticipated that when the quadrupole moment parameter is close to $1$,
the numerical solution obtained from the Hartle-Thorne gravity would be closer to the Kerr solution.
However, for $q=0$ and $q > 1$, the solutions obtained from Hartle-Thorne gravity models
deviate from the Kerr solution. Simultaneously, when comparing the results obtained from Hartle
-Thorne gravity with Schwarzschild, EGB, and Kerr, it can be stated that the disk density at
$r=4M$ takes similar values. The deviation occurs only in the cases of $q=0$ and $q>1$.

\begin{figure*}
  \vspace{1.4cm}
  \center
  \psfig{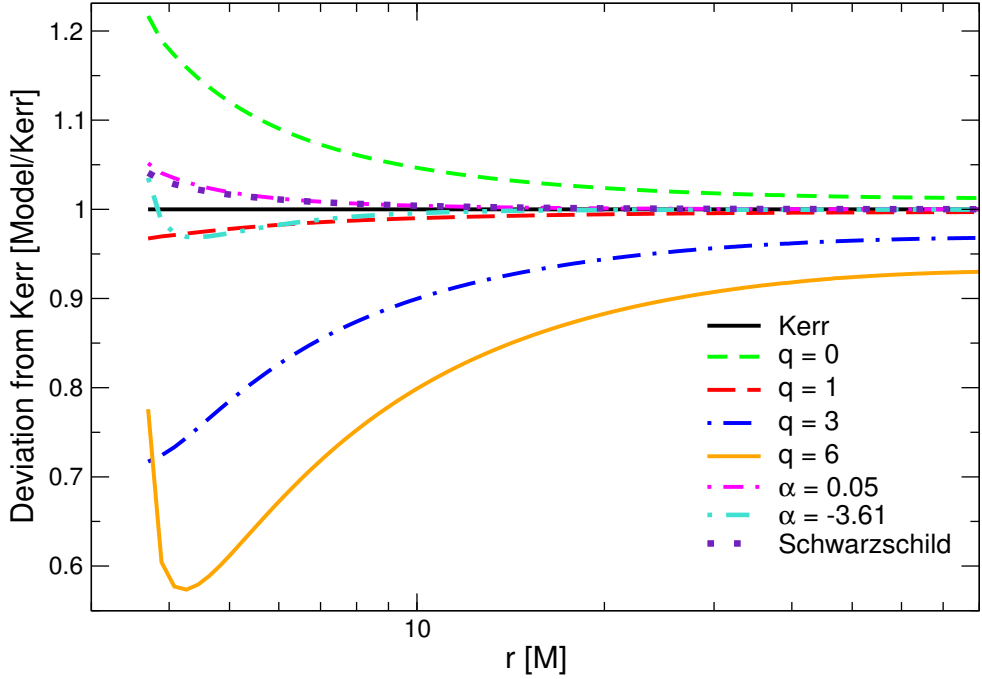}   
  \caption{The same as Fig.\ref{a028_deviation} but it is for a spin parameter of $a/M=0.9$.}
\label{a09_deviation}
\end{figure*}

For the rapidly rotating black hole model, the deviation ratio from the Kerr black hole is
presented in Fig.\ref{a09_deviation}. As observed, the minimum deviation from the Kerr solution
for our models is noted at $q=1$, in line with the previously mentioned observation that
the quadrupole moment parameter in the range of $0.5 < q < 1$ produces a solution similar to
Kerr. However, for $q=0$ and $q>1$, the solutions significantly diverge from Kerr solution.

On the other hand, in EGB gravity, the $\alpha=0.05$ provides a solution almost similar to
Schwarzschild. Generally, EGB gravity exhibits behavior similar to
Kerr for possible positive \cite{Donmez_EGB_Rot}
values of $\alpha$ and $alpha > -4$, as seen in the slow rotating black hole model.
Conversely, for extremely negative values of $\alpha$, as observed in this comparison,
the deviation from Kerr is substantial. Additionally, in this comparison, it is evident that
EGB gravity with large negative $\alpha$ values shows similar behavior to Hartle-Thorne with
large $q$, especially in the region near the black hole's horizon.
However, the deviation ratios from Kerr differ significantly between these two gravities
for bigger $q$. This similar behavior in both gravities can be utilized to explain emissions
occurring in strong gravitational fields and their physical mechanisms.

\begin{figure*}
  \vspace{1.4cm}
  \center
  \psfig{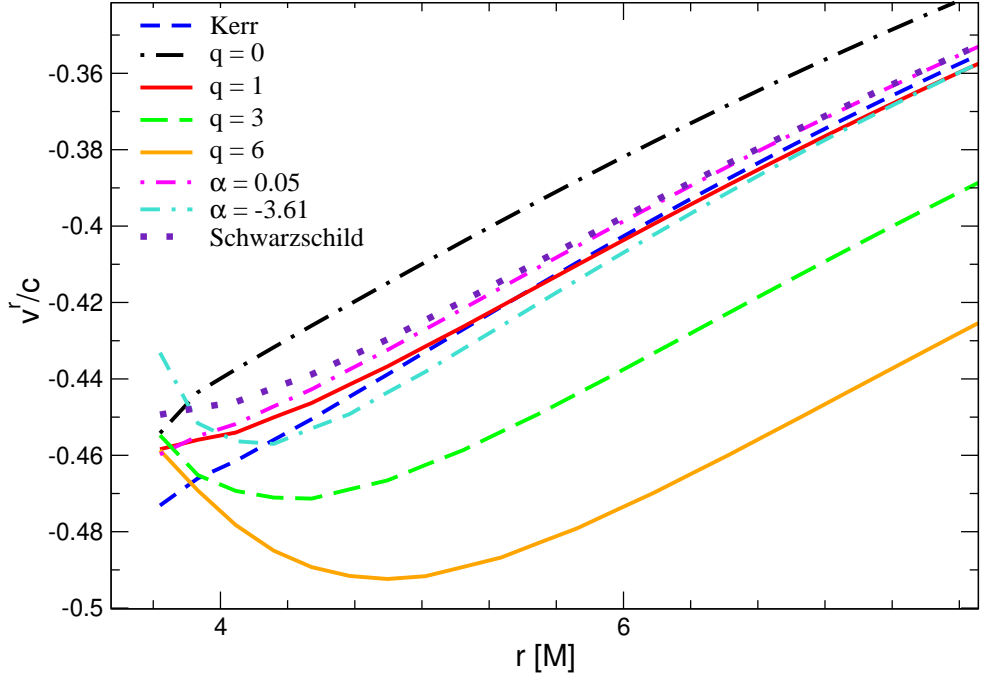}    
  \caption{The same as Fig.\ref{a028_mass_acc} only  for the  radial velocity of
   the matter falling toward the black hole but it is for a spin parameter of $a/M=0.9$.}
\label{a09_mass_acc}
\end{figure*}

Falling matter's radial velocity towards a black hole provides crucial information about the
structure of the accretion disk and the behavior of matter. The transformation of the disk's
angular momentum or the emission of generated energy can be deduced from it. Details about the amount
of matter residing on the disk, the ratio of matter falling toward the black hole, and whether
the disk is hot or cold can be inferred. Therefore, knowing the radial velocity is important when
comparing different gravities. This way, based on the characteristics of observed data, the effects
of gravity can be revealed, or the presence of a particular gravity can be identified.
In Fig.\ref{a09_mass_acc}, the variation of radial velocity with respect to $r$ for different
gravities is shown according to the parameters used in the models. For Hartle-Thorne gravity, again, the
best fit with Kerr is observed from the results obtained for the $q=1$ model. As previously
mentioned, other $q$ values deviate significantly from Kerr results, as clearly depicted in
the figure. Our comparisons and results from the previous paragraph are evident in this figure.
Particularly for $q > 1$ and large negative values of $\alpha$, it is observed that
the matter slows down as it approaches the black hole.

\subsection{Comparison of Slowly and Rapidly Spinning Black Holes in The Hartle-Thorne gravity}
\label{compare_result}

The Hartle-Thorne metric is a solution to the Einstein field equations for compact objects and
typically describes the spacetime around a slowly rotating Kerr black hole. It is a gravity
solution obtained by expanding the metric that defines the space for a fully rotating black hole,
taking into account higher-order terms in a series. It is generally applied in cases where the
angular momentum of the black hole is small.

\begin{figure*}
  \vspace{1.4cm}
  \center
  \psfig{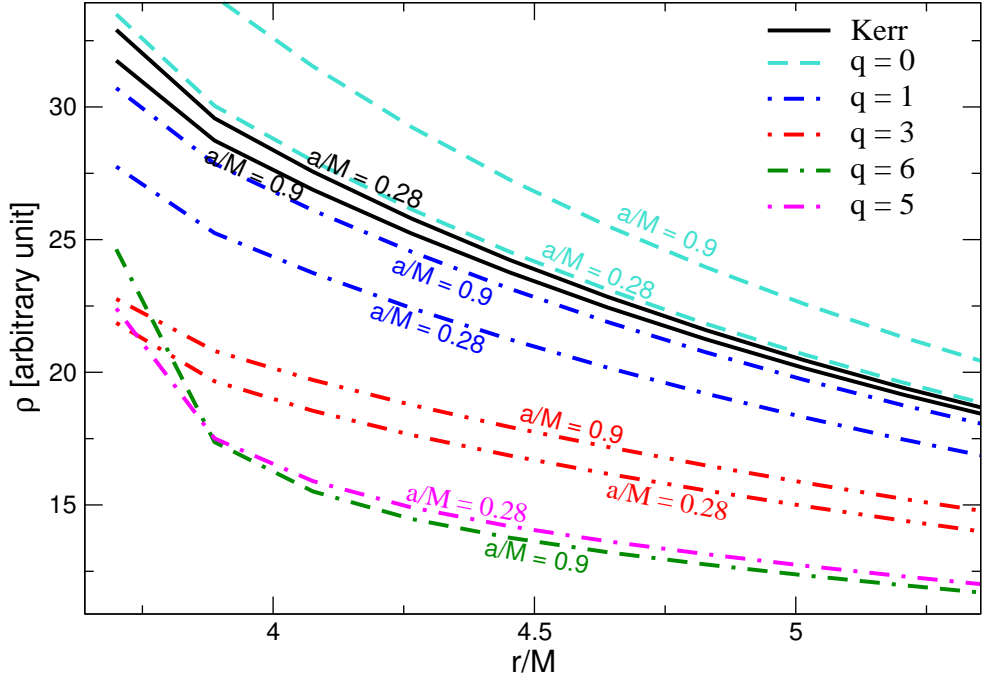}   
  \caption{Accretion disk density variation with respect to $r$ for different $a/M$ and $q$ values
    in Kerr and Hartle-Thorne gravities. The agreement and differences between the results
    obtained using the Kerr spacetime and the calculated density of the disk for different
    $q$ values have been highlighted. It is observed that the density of the disk undergoes
    significant changes with different values of $q$.}
\label{compare}
\end{figure*}

In this study, to demonstrate that the Hartle-Thorne metric can be used even in cases of
high rotation parameters of a black hole, a stable disk formed around the black hole is modeled.
In Fig.\ref{compare}, the radial variation of the disk's density, dependent on the quadrupole
moment $q$, is shown for a black hole model with a rotation parameter of $a/M=0.9$ using
the Hartle-Thorne metric and compared with the Kerr black hole solution.
As seen in Fig.\ref{compare}, for a slowly rotating black hole with $a/M=0.28$, the
$q=0$ provides a result close to the Kerr solution, while other $q$ values deviate from
the Kerr solution. On the other hand, for $a/M=0.9$, the solution obtained from Hartle-Thorne
gravity at $q=1$ is closer to the Kerr solution. However, for $q$ values other than $1$,
the deviation is significant and increases as the value of $q$ increases.

According to these initial results, the Hartle-Thorne metric can be used for the rapidly rotating
black holes, but $q$ should be different for a given specific $a/M$. If the
result obtained from the Hartle-Thorne gravity is desired to transform into the Kerr solution,
$q$ should increase as $a/M$ increases. On the other hand, the results obtained from
Hartle-Thorne gravity produce different results from Kerr for different values of $a/M$ and $q$.
For $q$ values other than $q=q_{Kerr}$, which are compatible with Kerr for Hartle-Thorne
gravity, either the density of the disk is greater or less than Kerr. In the case of
$q < q_{Kerr}$, the rest-density of the disk around the black hole increases, while in the
case of $q > q_{Kerr}$, this density decreases. This can alter the physical characteristics
of radiation in a regions with a strong gravitational field.

\section{Discussion and Conclusion}
\label{Conclusion}

In this study, we modeled the accretion disk around the black holes formed through the
spherical accretion by numerically solving the GRH equations using Schwarzschild, Kerr,
EGB, and Hartle-Thorne gravities. Our modeling is conducted in $2D$ on the equatorial plane.
When determining the model types, we consider possible different values of parameters
defining these gravities for both slowly rotating and rapidly rotating black holes.
Thus, we systematically conduct a study by considering the impact of $\alpha$ in EGB gravity
and the quadrupole moment in Hartle-Thorne gravity, along with the rotation parameter of
the black hole, on the formation and dynamic structure of the disk in different gravities.

To unveil the dynamic structure of the disk, we calculate the angular and radial variations
in disk density, the mass accretion rate of matter falling toward the black hole,
and the radial velocity of matter as it falls toward the black hole through spacetime.
By doing so, we reveal the dynamic structure of the disk and the behavior of matter around
the black hole that could shed light on observational data. Simultaneously, we compared the
outcomes of the disk's behavior in different gravities by normalizing them with Kerr.

According to the results from our numerical modeling, it has been observed that different
parameters defining gravity, namely $\alpha$ and $q$, play an effective role in the
formation of the disk around a slowly rotating black hole with $a/M=0.28$. The results
obtained from Hartle-Thorne are found to be consistent with Kerr for the range $0 < q < 1$.
Moreover, it is anticipated that the Hartle-Thorne solution could align with Kerr
when the value of  $q$ is around $0.5$
or closer to $q=0$. On the other hand, for  $q \ge 1$, it is noticed that the
results deviate from those obtained from Kerr. As  $q$ increases, this deviation is
observed to grow. In fact, when $q=5$, it is determined that the radial velocity of
the disk decreases near the black hole horizon. In other words, as $q$ increases,
there seems to be a significant change in the behavior of the parameters responsible for the
formation of the disk. Furthermore, a comparison is made between EGB gravity and Kerr for the
case $a/M=0.28$. Although the results in this gravity do not deviate from Kerr as much as
in the case of Hartle-Thorne, it is observed that near the black hole horizon,
the change in density at a large negative $\alpha$ value is similar to the behavior of
the large  $q$ value in Hartle-Thorne.

The Hartle-Thorne gravity is generally formulated to describe the spacetime around
slowly rotating black holes. However, in this study, we have extended our analysis by
comparing the results obtained from Hartle-Thorne for a rapidly rotating black hole
(specifically, $a/M=0.9$) with Kerr and EGB gravities. This comparison is motivated by
the belief that it may provide insights into certain astrophysical phenomena that remain
unexplained in black hole observations. The numerical simulations reveal that the disk's
behavior around rapidly rotating black holes yields a solution that is closest to Kerr when
$q=1$ in Hartle-Thorne gravity. In essence, this suggests compatibility between Kerr and
Hartle-Thorne solutions for values of $q$ slightly less than $1$. Conversely, for both
$q=0$ and $q>1$, it becomes evident that Hartle-Thorne deviates from the Kerr solution.
Analogous to the scenario of a slowly rotating black hole model, in Hartle-Thorne gravity with
$q=6$ and EGB gravity with $\alpha=-3.61$, despite both gravities exhibiting similar behavior
near the black hole's horizon, the EGB solution aligns much more closely with the Kerr solution.

In conclusion, the results obtained in this study and the comparison of the Hartle-Thorne solution
with various gravities, particularly for slow and rapidly rotating black hole models, can be
further expanded to encompass wider ranges of $q$. This extension would facilitate the
identification of the specific values of $q$ at which the Hartle-Thorne solution undergoes
a transformation into the Kerr solution. A more detailed application of Hartle-Thorne gravity
to black hole models could significantly contribute to elucidating certain astrophysical
systems characterized by unclear physical mechanisms
\cite{Andersson2001CQGra,Stergioulas2003LRR}.


\section*{Acknowledgments}
All simulations were performed using the Phoenix  High
Performance Computing facility at the American University of the Middle East
(AUM), Kuwait.\\

\section*{Data availability}
The data used in the article are entirely composed of data obtained by running our own program,
which solves the General  Relativistic Hydrodynamics Equations, on a High-Performance Computer.
Each model has generated about 4GB of data. Since the total number of
models is 19 given in Table \ref{Inital_Con}, this means a
total of 36GB of data. Data can be sent to the requested people through private communication.

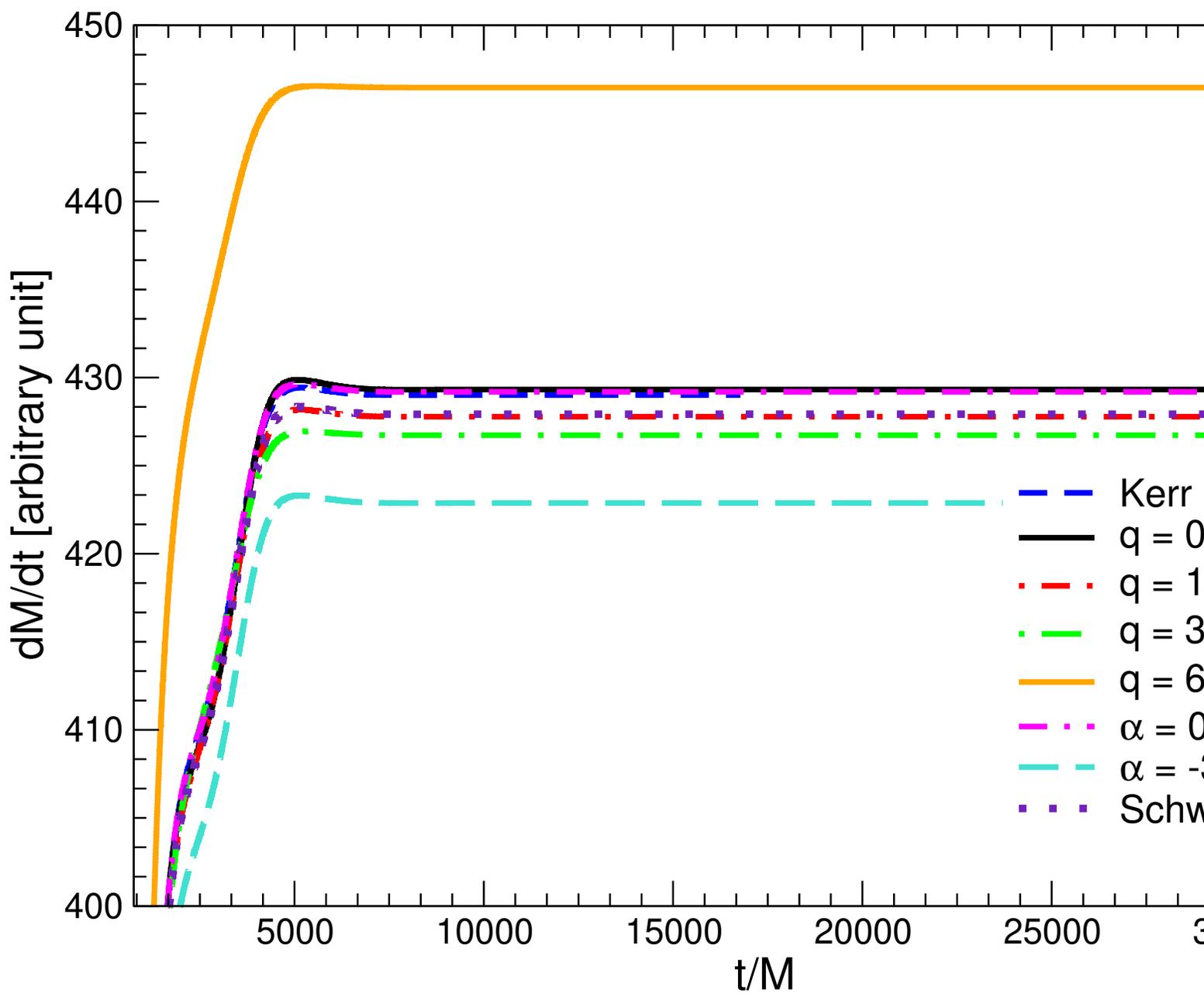
\end{document}